\documentclass[10pt,twocolumn]{article}
\usepackage[letterpaper,margin=0.72in,columnsep=0.25in]{geometry}
\usepackage[T1]{fontenc}
\usepackage{lmodern}
\usepackage{microtype}
\usepackage{amsmath,amssymb}
\usepackage{graphicx,booktabs,tabularx,array}
\usepackage[font=small,labelfont=bf]{caption}
\usepackage[authoryear,round]{natbib}
\usepackage{url}
\usepackage[breaklinks=true,hidelinks]{hyperref}
\usepackage{titlesec}
\usepackage{fancyhdr}
\usepackage{enumitem}
\titleformat{\section}{\large\bfseries}{\thesection}{0.6em}{}
\titleformat{\subsection}{\normalsize\bfseries}{\thesubsection}{0.6em}{}
\titlespacing*{\section}{0pt}{2.2ex plus .7ex}{1ex}
\titlespacing*{\subsection}{0pt}{1.8ex plus .5ex}{0.6ex}
\setlist{nosep,leftmargin=*}
\providecommand{\tightlist}{\setlength{\itemsep}{0pt}\setlength{\parskip}{0pt}}

\hypersetup{pdftitle={Seal, Then Sample: Sampled Layerwise Proofs for
Verifiable LLM Inference from GPT-2 to
70B},pdfsubject={Sampled Layerwise Proofs for Verifiable LLM Inference}}
\begin{document}
\twocolumn[{
\centering
{\LARGE\bfseries Seal, Then Sample: Sampled Layerwise Proofs for
Verifiable LLM Inference from GPT-2 to 70B\par}
\vspace{0.75em}
{\normalsize Youki Lim, Sam Yong\par}
\vspace{0.25em}{\small TrueOpen\par}
\vspace{0.5em}
{\small 22 September 2026\par}
\vspace{1em}
\begin{minipage}{0.94\textwidth}
\small
\noindent\textbf{Abstract.} Verifying outsourced language-model
inference requires a precisely identified computation and an audit whose
cost a service can afford. We present Sampled Layerwise Proofs (SLP), a
protocol and prototype that commits the boundary activations of every
chunk of an inference trace, absorbs all commitments before any
challenge is drawn, and then proves a verifier-selected subset of chunks
together with the chunks that bind the prompt and the answer. Audit
coverage becomes a runtime parameter over one set of commitments: on a
TinyLlama-1.1B trace, proving seven of 47 chunks takes 22.0\% of the
time and 6.8\% of the proof size of proving all 47. Because proof cost
is dominated by weights rather than tokens, SLP packs concurrent
requests into one trace under a block-diagonal causal mask and binds
each request's prompt and answer to its slot. Twelve packed requests are
proved in 181.9 s, 6.5 times less than twelve separate proofs at the
measured single-proof cost, and a simulated service proves twelve
requests at 30.6 s per request with 0.6 s of verification each,
rejecting a tampered answer. Disk-backed integer weights and streamed
polynomial commitments let a single Llama-2-70B run complete on a 2 TB
CPU host: 163 chunks sealed, five proved, a 4.34 MiB proof in 1,259 s,
verified in 46.3 s without the weights. The proven object is a
fixed-point canonical model; we trace a severe fidelity loss to the
residual-stream bit width, repair it with an LLM-aware observer, and
measure 84.8-84.9\% argmax agreement with the floating-point reference
over 334,705 WikiText-2 test positions. The limits are stated as
precisely: guarantees cover proven chunks only, a fixed invalid chunk in
the 70B setting is covered with probability 3/161, a manifest-only
Fiat-Shamir schedule can be ground at 12.5 ms per attempt and needs an
externally ordered challenge, and all measurements use a test reference
string.
\par\vspace{0.6em}
\end{minipage}
\vspace{1.3em}
}]
\section{Introduction}\label{introduction}

A language-model service returns tokens, but the tokens alone do not
identify the computation that produced them. A provider may use
different weights, lower precision, or fewer operations than the
customer requested. Re-executing the entire inference defeats much of
the purpose of outsourcing. Cryptographic verification offers another
route: identify a model and an input, express the computation as a
relation, and check a proof of that relation. For large transformers the
proof workload, and the memory that its representations require, exceed
those of ordinary inference by orders of magnitude.

This paper makes the amount of computation proved an explicit,
adjustable quantity. Sampled Layerwise Proofs (SLP) first commits the
activations at every boundary of a partitioned inference trace and
absorbs every commitment into the transcript. Only then does it derive
which chunks to prove, and it always proves the chunks adjoining the
input and the output. The verifier checks every boundary connection,
verifies the arithmetic inside the selected chunks, and can later
request any further chunk over the same commitments, up to all of them.
Increasing the sample increases coverage; selecting every candidate
recovers full chunk coverage.

The distinction between commitment coverage and arithmetic coverage runs
through the paper. A complete manifest prevents a shared boundary from
being changed after it is fixed, and a valid proof of the final
projection establishes how its committed input relates to the answer.
Neither establishes that an unproved interior transition is correct. In
the 70B configuration evaluated here, three random candidates and two
anchors are proved out of 163 chunks, so a fixed single invalid
candidate is covered with probability \(3/161\) under uniform sampling.
What SLP contributes is a configurable audit interface with a measured
cost curve, and a systems realization that reaches 70B; it is not a
stronger detection law for a raw committed trace.

Proof cost in this setting is dominated by model weights, not by tokens:
on TinyLlama, doubling the context from 8 to 16 changed the proving time
from 94.8 s to 99.0 s. That observation motivates the second half of the
design. SLP packs several concurrent requests into one trace with a
block-diagonal causal mask, decodes the slots in lockstep, and binds
each request's prompt and claimed answer to its slot inside the
verification interface. One proof then covers many requests, and the
per-request proving cost falls with the batch size. We show the
amortization on twelve concurrent requests, verify that the mask leaks
nothing between slots, and connect the packed proof to a simulated
service in which answers are returned before proofs and a tampered
answer is rejected. Packing also changes the security accounting,
because one sampled set now audits many requests; we make that
accounting explicit.

SLP builds on deep-prove's chunked sumcheck and polynomial-commitment
implementation \citep{deepprove2026}. Prior work established full
transformer proofs \citep{zkllm2024}, layerwise audit budgets
\citep{nanozk2026}, and decomposed proofs that share boundary
commitments \citep{zkcomposer2026}. SLP combines a sampled proving path
with mandatory endpoint binding, a separate sealing interface, packed
inference traces, and streamed model registration, and measures the
resulting costs and limits across GPT-2, TinyLlama-1.1B, and one
Llama-2-70B run.

Two engineering problems shape the systems half. The first is
representation size. With eight bytes per integer weight and 32 bytes
per field element, 70 billion parameters occupy about 560 GB and 2.24 TB
respectively, and holding both representations would need about 2.8 TB.
We spill quantized weight tensors to disk and commit weight polynomials
in bounded batches. The recorded 70B loading and commitment stages
peaked at about 387 GB resident memory and a 60 GB working set.
Inference still materializes model-sized integer state, so the complete
pipeline is not yet single-layer-memory; Section 6.4 reports the stages
separately.

The second problem is what a successful proof says about the advertised
model. The backend runs a fixed-point canonical model derived from a
floating-point checkpoint, and a commitment to that model cannot certify
that it behaves like its parent. In early TinyLlama experiments the
generic quantization observer compressed the residual stream to 12 bits
and the model produced nonsense. Switching to an LLM-aware observer that
keeps the residual stream at about 24 bits restored agreement with the
floating-point model, with the nominal linear-input bit width unchanged.
We evaluate the repair with short greedy continuations and with a paired
evaluation over the WikiText-2 test split, and we treat fidelity as a
reported property of a registered model version, separate from the
soundness of its proof.

A correct implementation of deterministic sampling leaves one protocol
gap that we measure rather than assume. If the challenge is derived only
from a worker-chosen manifest, the worker can fabricate a boundary,
seal, inspect the sample, and repeat until the sample avoids the
fabricated chunks. On GPT-2 one attempt costs 12.5 ms. The library
therefore separates sealing from proving and accepts an external
challenge; we specify the ordering, uniqueness, and authentication that
a service must provide around it, and we distinguish that adaptive audit
model from the beacon-less configuration used for the performance
measurements.

The contributions are:

\begin{enumerate}
\def\labelenumi{\arabic{enumi}.}
\tightlist
\item
  \textbf{A seal-then-sample protocol with mandatory endpoint binding}
  (Section 3): complete boundary manifests, a verifier-owned sampling
  policy, anchors derived from the plan, six verification checks, a
  coverage bound with its assumptions, and an escalation path to full
  coverage over the same commitments.
\item
  \textbf{Packed batch proving with per-slot binding} (Section 4): a
  block-diagonal mask, lockstep decoding, per-request prompt and answer
  checks, a measured non-interference test, and the security accounting
  of a shared sample; twelve requests in one proof at 6.5 times less
  proving time than separate proofs at the measured single-proof cost.
\item
  \textbf{Streamed registration at 70B} (Section 5): disk-backed integer
  weights and bounded-batch weight commitments that complete a
  Llama-2-70B sealing, sampled proof, and verification on a 2 TB host.
\item
  \textbf{Same-trace cost measurements} (Section 6): five audit
  strengths on one TinyLlama trace, a complete 70B run with per-stage
  accounting, packing and service experiments, and a measured grinding
  attack with its beacon control.
\item
  \textbf{A diagnosis and paired evaluation of canonical-model fidelity}
  (Section 7): the residual-stream root cause, the observer repair, and
  84.8-84.9\% argmax agreement over 334,705 positions.
\end{enumerate}

We make no priority claim for layerwise commitments, no zero-knowledge
claim for this implementation, and no claim that low-coverage proofs
authenticate an entire inference. Section 8 collects the limitations in
one place.

\section{Computation, Commitments, and
Scope}\label{computation-commitments-and-scope}

\subsection{The object being verified}\label{the-object-being-verified}

Let \(v\) identify the registered model configuration and let \(x\) and
\(y\) be the public prompt and claimed continuation. The inference trace
is partitioned into \(N\) chunks according to a plan derived from the
verification context. For chunk \(i\), let \(u_i\) collect its boundary
inputs and outputs, \(w_i\) its registered weights, and \(z_i\) its
internal witness. The backend checks a relation
\(R_{v,i}(u_i,w_i,z_i)\), and a sound proof establishes the existence of
an admissible witness for that encoded relation. Identifying the
relation with the intended deterministic model function requires the
rounding, range, normalization, and token-selection rules to be
constrained as intended.

Two obligations are therefore separate. The encoding obligation asks
whether the circuit captures the desired model semantics; proof-system
soundness asks whether an adversary can prove a false circuit statement.
Tests that accept honestly generated traces show functional
compatibility and settle neither question for an arbitrary malicious
witness. The conditional guarantees in Section 3 are stated for the
backend's encoded relation.

The encoding obligation is concrete in the current code. The LayerNorm
verifier constrains both a sum and a magnitude error bound, and the
RMSNorm verifier a magnitude error bound; both use the generic
normalization lookup verifier. The resulting finite-field relation
admits an approximation range, and we have not proved that every witness
inside that range yields the same output tokens. Our coverage analysis
therefore counts relations that are false under the implemented
constraints, and Section 8.2 discusses what this leaves open.

The canonical configuration consists of quantized linear weights,
activation scales, model structure, tokenizer behavior, context and
position conventions, and greedy token selection. The implementation
exposes SHA-256 digests over the linear weights and the
activation-scaling metadata, which identify changes to the artifacts
they cover. A registry must additionally bind normalization parameters,
graph interpretation, tokenizer revision, decoding policy, and backend
version to the verification context; the current receipt is a compact
audit handle for such a registry rather than the registry itself.

The floating-point reference is a separate object. The fidelity
experiments compare the canonical runtime with an f32 execution in the
same framework under a specified input construction. They measure
agreement with that reference, not equality with an arbitrary fp16
service, another inference framework, or another quantization scheme,
and a proof of canonical replay is not evidence that an earlier native
floating-point execution took place.

\subsection{Backend and
representation}\label{backend-and-representation}

The implementation uses the model graph, chunk prover, and verifier from
deep-prove. Linear layers are reduced by sumcheck in the GKR style
\citep{gkr2015, thaler2013}; nonlinear operations use lookup and
associated arithmetic constraints in the LogUp family \citep{logup2022}.
The polynomial commitment is HyperKZG over BN254 and the transcript uses
Blake3. Model-weight commitments are created once at registration and
reused across requests; a chunk proof opens the boundary and weight
polynomials at the points its reduction requires.

KZG-style commitments rest on a structured reference string
\citep{kzg2010}. In this prototype the commitment context is created
with the backend's \texttt{test\_setup} routine, which generates the
parameters locally. Such parameters are adequate for measuring cost, and
they are not a secure setup against the party that generated them, who
may know the trapdoor. A deployment needs parameters of the correct
curve, format, and size from an authenticated ceremony, together with
authenticated verification contexts. All performance results in this
paper were obtained under test parameters.

The prototype adds no masking and makes no zero-knowledge claim. Public
inputs and outputs are visible to the verifier, and the unmasked
argument may reveal information about witness or weight polynomials. We
use the established term zkML for the research area and the underlying
systems, and describe SLP's output as sampled arithmetic proofs.

\subsection{Actors and adversary}\label{actors-and-adversary}

A registrar associates a named model version with a verification context
and supporting metadata. A worker executes the model and produces the
trace and proofs. An auditor holds an authenticated copy of the context
and fixes the partition and the sampling policy. The roles may be held
by different parties; if the worker is also the registrar, an external
model-identity check is still needed, because a commitment binds
particular weights, not a marketing name.

The worker may fabricate intermediate states, skip computation, change
precision, return a different answer, or try to influence the selection
of chunks. It controls all prover-side memory and may choose its
response after seeing the prompt. In the adaptive audit model it cannot
change the auditor's context, modify an accepted manifest, or predict or
bias the challenge before the manifest is fixed. It may still refuse to
answer after seeing an unfavorable challenge; refusal is a service
outcome that an external policy must treat separately from acceptance.

The analysis assumes binding commitments, sound backend arguments,
correct configuration authentication, and adequate transcript domain
separation. We do not prove the underlying primitives secure, and our
code review covers the sampling, manifest, beacon, chunking, setup, and
normalization paths rather than every operator. The test reference
string is deliberately outside these assumptions for the benchmark
configuration.

\subsection{Audit strengths}\label{audit-strengths}

The library supports manifest checks and sampled chunk proofs with the
sample count clamped to the available candidates. With zero random
candidates, the input and output anchors are still proved; with all
candidates, every chunk relation is checked, which is full chunk
coverage conditional on the backend and the encoded relation. Neither
setting is a proof of the amount of physical computation performed.

Freivalds checks provide a separate lightweight diagnostic for matrix
multiplication \citep{freivalds1977}. For \(Y=WX\), an auditor with
access to the operands compares a random projection of each side, and
the usual algebraic error bound applies to a fixed incorrect product
over a field. A commitment alone does not let the auditor project secret
weights; authenticated openings, the weights themselves, or another
proof mechanism are required. The standalone field and floating-point
examples in this project are diagnostics and are not yet an integrated,
commitment-bound audit tier for LLM matrix products.

\section{Seal-Then-Sample Protocol}\label{seal-then-sample-protocol}

\subsection{Registration and complete
manifests}\label{registration-and-complete-manifests}

Registration produces the model's verification context and the
commitments to its weight polynomials. For each request the worker runs
the canonical runtime, obtains the inference trace, and constructs from
the agreed graph and chunking policy a complete boundary manifest: the
input and output ports of every chunk, the polynomial commitment to each
boundary tensor, and the shapes and graph locations to which the values
belong. Adjacent chunks share one committed boundary value rather than
holding separate local copies.

Let \(A\) be the set of chunks touching public input or output and \(C\)
the remaining candidates, with \(M=|C|\). The random sample size is
\(s\leq M\), and the prover must supply proofs for exactly the union of
\(A\) and the selected candidates. The anchors are derived from the
plan, never chosen by the worker; the measured decoder configurations
have two. We count anchors separately whenever we quote coverage.

The chunking code derives the plan from the model and its constraints.
For the LLM configurations measured here the count is \(2L+3\) for \(L\)
transformer blocks: one split after every residual addition gives
\(2L+1\) interior segments, plus a standalone embedding chunk and a
standalone final-projection chunk, giving 47 chunks for TinyLlama and
163 for Llama-2-70B. Horizontal partitioning of the projection can
change the constant, so reproduction uses the recorded plan rather than
the formula.

Sealing is a commitment step, not a correctness proof. A manifest can be
structurally consistent while containing values that no honest execution
would produce; the arithmetic proofs of the selected chunks are what
rule this out, chunk by chunk. This fixes the unit of analysis: an
invalid chunk is a false encoded boundary-to-boundary relation, which
need not coincide with a skipped layer, a modified tensor element, or a
unit of saved compute.

\subsection{Challenge generation and external
ordering}\label{challenge-generation-and-external-ordering}

The transcript is initialized from the verification context and absorbs
the boundary manifest before any candidate identifier is derived. Public
IO is bound through the anchor proofs rather than absorbed separately.
The implementation can additionally absorb externally provided beacon
bytes. Each draw appends a counter to the transcript input, so repeated
extraction from an unchanged state cannot occur, and candidates are
collected without replacement. The helper maps the low 64 bits of a
field challenge modulo the candidate count and fails closed if more than
\(64(M+16)\) draws would be needed. The equations below describe ideal
uniform sampling; an implementation-level argument must also account for
the modular mapping and the draw cap.

Fiat-Shamir makes an interactive challenge reproducible from a
transcript \citep{fiatshamir1986}; it does not make a worker-chosen
transcript unpredictable to that worker. Under a manifest-only schedule
an adversary can fabricate a boundary, seal, inspect the selected set,
and repeat. Commitment binding prevents changing an accepted commitment;
it does not prevent this offline search over candidate commitments.
Section 6.6 measures the search.

For an adaptive audit, SLP therefore requires an external ordering. The
auditor assigns a session identifier that binds the model context, the
request, the claimed answer, the policy, and the packed layout where
applicable. An authenticated service records exactly one accepted
manifest for that session before any challenge is available. The auditor
then supplies a fresh nonce, or the parties take the value of a
predetermined public beacon round. The sample is derived from the
accepted manifest, the challenge, and the fixed policy, and the auditor
checks the recorded ordering with the expected challenge, independently
of the worker.

Uniqueness must hold at the request level. Registering many candidate
manifests before a beacon and choosing one afterwards preserves the
grinding advantage, as does letting the worker choose which past beacon
round to use. Retries and alternative session identifiers must not
create fresh opportunities for the same obligation. A service can reject
duplicate manifests, bind retries to the original request, and record
missed proof deadlines; what penalty follows a refusal is a deployment
decision.

The library implements the seal/prove separation and the expected-beacon
verification interface. The authenticated registration service, durable
uniqueness enforcement, and a production beacon integration are external
obligations; Figure 1 separates the two. A nonce exchanged interactively
also needs authenticated evidence of the exchange if a third party later
verifies the result, and a timestamp attached only to the beacon does
not establish when the worker's manifest became binding.

\begin{figure*}[t]
\centering
\includegraphics[width=\textwidth]{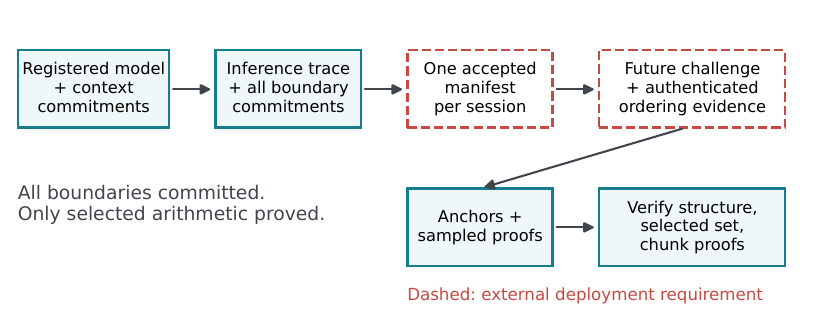}
\caption{SLP audit ordering. Solid boxes are the library's data and proof operations. Dashed boxes are the registration and challenge obligations of the adaptive audit model, which a deployment provides around the library; the benchmark configuration runs without them.}
\end{figure*}

\subsection{What acceptance
establishes}\label{what-acceptance-establishes}

For a fixed accepted manifest, let \(B\subseteq C\) be the candidates
whose committed boundaries violate the encoded chunk relation, with
\(b=|B|\). Assume any invalid anchor is rejected, boundary and weight
commitments are binding, and a false selected relation is accepted with
total probability at most \(\epsilon_{\mathrm{backend}}\). Under uniform
sampling without replacement,

\[
q(M,b,s)=\frac{\binom{M-b}{s}}{\binom{M}{s}},
\qquad
p(M,b,s)=1-q(M,b,s).
\]

Here \(q\) is the probability of selecting no invalid candidate; if
\(s>M-b\) the numerator is zero. An invalid anchor is always checked, so
the sampling term concerns candidates only. For invalid candidates, the
probability of accepting the false manifest is at most
\(q+\epsilon_{\mathrm{backend}}\) plus any deviation from ideal
sampling, and the rejection probability is at least
\(p-\epsilon_{\mathrm{backend}}\) minus that deviation. These are
conditional bounds on encoded relations.

The argument conditions on the sample. A sample that intersects \(B\)
requires acceptance of at least one false selected statement; a sample
that misses \(B\) can consist entirely of valid statements although the
trace is invalid. The construction contains no error-correcting encoding
of the trace, so there is no distance amplification that turns a single
invalid candidate into many sampled inconsistencies, and strengthening
the per-chunk proof does not change the probability that the invalid
chunk is never selected.

For the recorded 70B plan, \(N=163\), \(|A|=2\), and \(M=161\). The
measured sample count is \(s=3\), so a single invalid candidate is
covered with probability \(3/161=1.863\%\); five is the number of proved
chunks, not the sample count. At \(s=20\) the same coverage would be
12.42\%, and for \(b=8\) and \(s=20\) about 66.3\%. Fabricating one
boundary between two otherwise honest chunks invalidates two relations,
one on each side, and an attacker who wants exactly one invalid relation
must recompute a consistent downstream suffix. We keep \(b=1\) as the
worst-coverage case for the relation-level analysis without claiming
that every answer substitution can be realized at the cost of one
boundary.

Under \(k\) independent, enforced audits with conditional detection
probability \(p\), cumulative detection is \(1-(1-p)^k\). At
\(p=3/161\), 37 audits reach 50\% and 160 reach 95\%. The count refers
to independently sampled audits of the same corruption; a single request
that is never audited again gets one draw, and the requests inside a
packed proof share one draw (Section 4.4).

\subsection{Anchors and verification}\label{anchors-and-verification}

The input anchor connects the public tokens to the first committed
boundary, and the output anchor connects the last committed boundary to
the claimed token selection. Anchoring keeps a proof from floating free
of its request and answer, and it rejects an answer changed after the
proof was fixed. It does not make the interior arithmetic correct: an
attacker can compute an honest output anchor on a fabricated interior
boundary. Changing a served answer and constructing a different pre-seal
trace are different attacks with different coverage.

Verification proceeds in six checks. The verifier (1) derives the
expected plan from its own context and checks that it is a complete
partition; (2) checks the manifest's coverage and shared-port
consistency and replays the required shape and quantization metadata;
(3) derives the anchors and candidates and reproduces the sample from
its own policy and challenge; (4) requires the supplied chunk proofs to
equal the expected set exactly; (5) verifies each selected chunk against
the manifest's committed values, comparing commitment bytes under the
binding assumption; and (6) checks the anchor relations against the
public prompt and answer. A receipt may summarize counts and IO digests,
and the verifier never takes policy parameters from it.

A mismatched beacon generally changes the selected set and is rejected
at check (4); the exact behavior depends on the complete transcript
binding, and we report the mismatch tests as functional negatives.
Verification without weights is not verification in constant time: the
verifier reconstructs the model-wide partition and boundary structure
before checking the selected arithmetic, so its cost has a model-wide
structural component and a selected-proof component. We report the
combined time.

\subsection{Full coverage and
escalation}\label{full-coverage-and-escalation}

Selecting every candidate proves every chunk over the sealed boundaries;
the TinyLlama experiment in Section 6.2 completes and verifies this
path. A later full-coverage audit reuses the model commitments and the
sealed boundaries but needs a new proof object under the new policy, so
witness reconstruction, sealing, serialization, and communication recur.
The worker must also have retained the trace. The current system has no
deadline-enforced dispute service or retention contract, so we describe
escalation as an available relation and software path that a service
must arrange to execute. The 70B all-candidate run was not performed.

\section{Packed Batch Proving}\label{packed-batch-proving}

\subsection{Layout and block-diagonal
mask}\label{layout-and-block-diagonal-mask}

Proof cost in this backend is dominated by the weight-dependent work of
each chunk, and the context-length experiment in Section 6.5 shows that
doubling the context changed the proving time by 4.4\%. Packing exploits
this. Several requests are assigned to equal-length slots of one larger
trace, and a block-diagonal causal mask lets each query attend only to
keys inside its own slot while preserving ordinary causal order within
the slot. The layout fixes slot offsets, capacities, and prompt lengths.
The prover and the verifier evaluate the same mask, including the
verifier's closed-form treatment of its multilinear representation.
Greedy decoding advances all slots in lockstep, and each prompt plus
continuation must fit its slot.

\subsection{Per-slot binding and
verification}\label{per-slot-binding-and-verification}

The packed proof is checked against every request it carries. The
verification interface receives each request's prompt and claimed answer
together with the packed proof, and it checks that the committed token
tensor contains the prompt at the slot's offset and the claimed
continuation at the following positions. The anchor relation and these
per-slot checks are both necessary: a structurally valid packed
computation alone does not tell a consumer which returned text belongs
to which request. The layout is therefore part of the computation
configuration that registration must fix.

\subsection{Position semantics and
non-interference}\label{position-semantics-and-non-interference}

Two functional tests characterize the packed runtime. The
non-interference test replaces the contents of all neighboring slots
while holding one slot fixed; the fixed slot's outputs are unchanged
token for token, so the mask leaks nothing across slots. The equivalence
test compares each slot with a standalone execution of the same prompt:
slot zero matches token for token, while slots at nonzero offsets can
differ, because rotary position encoding and fixed-point rounding use
the packed absolute positions. The packed runtime is thus a specified,
layout-dependent computation. Its fidelity to the floating-point model
at nonzero offsets has to be measured for the registered layout and
cannot be copied from the standalone evaluation (Section 7.5).

\subsection{Security accounting of a shared
sample}\label{security-accounting-of-a-shared-sample}

Packing amortizes weight-dominated proof work and changes the security
accounting. A batch with one sampled set has one set of audit challenges
rather than an independent draw per request. The probability of hitting
an invalid chunk relation is determined by the batch's invalid relations
and candidate set, and the cumulative-detection count of Section 3.3
applies to independent audits, not to the number of requests in a batch.
An economic argument that multiplies the packed request count by a
per-request detection probability overstates the opportunities to catch
the worker.

\section{Streamed Registration and Integer
Runtime}\label{streamed-registration-and-integer-runtime}

\subsection{Weight residency and streamed
registration}\label{weight-residency-and-streamed-registration}

The first memory bottleneck precedes polynomial commitments. At eight
bytes per weight, the quantized 70B weights occupy about 560 GB, and our
first attempt on a 128-core host with 503 GB of memory failed during
loading before any field representation existed. Memory-mapping the
checkpoint avoids one file copy but leaves the integer tensors that
quantization produces. We attach a persistent tensor store and spill
each quantized tensor to it, keeping a handle that describes how to
reload it. Resident and persistent paths are selected explicitly, so
smaller workloads keep their previous execution mode.

The second bottleneck is the weight representation for commitments.
Commitment inputs may now refer to disk-backed integer data instead of
resident field polynomials. The commitment loop loads one integer
tensor, converts it to field elements, computes the commitment, retains
only the small commitment and reconstruction metadata, and releases the
temporaries. Bounded batches allow commitment parallelism without ever
holding the whole model in field form; the batch and cache sizes are
memory configuration.

The savings are stage-specific. The recorded 70B loading and
quantization peak was about 387 GB, which still includes transient model
state, and weight commitment ran in about 60 GB of working memory. Both
registration stages therefore fit the 2 TB host with large margin.
Inference reloads the integer weights and does not yet release each
tensor after its last use, so model-sized integer residency remains a
property of the serving path, and a full prover could in principle
rebuild and release chunks sequentially as well. Sampling reduces the
arithmetic and reconstruction performed; bounded reconstruction reduces
simultaneous residency. Synthetic resident-versus-spilled tests confirm
that the two storage paths produce identical commitments and successful
openings.

\subsection{Integer inference and operator
costs}\label{integer-inference-and-operator-costs}

The canonical runtime stores many tensors as 64-bit integers although
their value ranges are far narrower. Separating storage width from
accumulator width would cut memory by four to eight times, and the
current implementation does not do it. Under the present representation
a 7B model needs about 56 GB for weights alone and does not fit the 24
GB GPU used here, and a quantized 70B model is larger in memory than its
fp16 parent. This is why our GPU measurements stop at 1.1B.

The GPU path executes integer inference on the tensor backend. Its
original lookup implementation incurred host-device transfers; moving
the lookup gathers onto the device removed them. The fidelity repair of
Section 7 widened the residual stream and exposed a second cost:
requantization rebuilt and transferred a \(2^{24}\)-entry table, 128 MiB
per column, on every call. Requantization and clipping are now evaluated
as device-side clamps with identical elementwise behavior in the tested
domain, and the smaller nonlinear tables are cached on the device. After
the repair, a TinyLlama forward pass took about 1.0-1.4 s at context 32,
4.0 s at context 128, and 26 s at context 512 on the recorded GPU;
context 1024 exhausted device memory. These are short profiling runs on
one device.

\section{Experimental Evaluation}\label{experimental-evaluation}

\subsection{Platforms, configurations, and evidence
status}\label{platforms-configurations-and-evidence-status}

We evaluate GPT-2, TinyLlama-1.1B-Chat-v1.0 \citep{tinyllama2024}, and a
Llama-2-70B checkpoint \citep{llama22023}. GPT-2 is loaded from a Q8\_0
GGUF file and the two Llama models from safetensors, so the three points
are separate measurements rather than one controlled scaling curve.
Table 1 summarizes the platforms.

\begin{table*}[t]
\centering
\small
\setlength{\tabcolsep}{3.5pt}
\renewcommand{\arraystretch}{1.18}
\begin{tabularx}{\textwidth}{>{\raggedright\arraybackslash}X>{\raggedright\arraybackslash}X>{\raggedright\arraybackslash}X>{\raggedright\arraybackslash}X}
\toprule
Workload & Model / chunks & Compute platform & Notes \\ 
\midrule
GPT-2 and synthetic tests & 124M, 27 chunks & Apple M3 Pro, 18 GB unified memory; CPU proofs & Historical OS/toolchain snapshot incomplete \\ 
TinyLlama inference and calibration & 1.1B, 47 chunks & NVIDIA RTX 4090, 24 GB; Ubuntu 24.04.4; CUDA 12.8 & Driver 595.71.05 \\ 
TinyLlama proofs and verification & Same traces & Host: two AMD EPYC 7742 CPUs, 128 physical cores / 256 hardware threads,
about 1,007 GB RAM & Prover parallelism varies by operation \\ 
Llama-2-70B pipeline & 70B, 163 chunks & Separate CPU host, 256 hardware threads, 2 TB RAM, 1.6 TB local disk & CPU model and software snapshot not recorded \\ 
\bottomrule
\end{tabularx}
\caption{Evaluation platforms. The GPU runs inference and calibration only; all
proofs and verifications in this paper were produced on CPUs.}
\end{table*}

All proofs use BN254/HyperKZG with locally generated test parameters and
the manifest-only, beacon-less challenge schedule; the beacon controls
are reported in Section 6.6. Builds use Rust's release profile, and the
GPU inference builds enable the CUDA backend. Archived logs show the
sumcheck backend reducing a requested thread count of 30 to 16, so the
proving times should not be read as full use of the 256 host threads.
The source workspace is a modified deep-prove checkout based on upstream
revision \texttt{9d1a53e2ef49ffa2c902b8689cd3c58057a4e662} with our
sampling, persistent-weight, quantization, lookup, and packing changes;
the revision material records source locations and log provenance.

We classify quantities as measured point values, arithmetic derived from
them, or unmeasured projections, and label each table accordingly. Most
timings are single observations. The 70B execution is one run. The
grinding experiment has 200 trials per setting, and the
repeated-inference check has 10 runs and 20 comparisons. Section 8.3
states the resulting limits on interpretation once, rather than after
every table.

\subsection{Same-trace full and sampled
proving}\label{same-trace-full-and-sampled-proving}

The smallest end-to-end configuration is GPT-2 on the laptop,
re-measured with the current build: inference 6.71 s; sealing 27
boundaries and proving five chunks (two anchors plus three samples) 28.0
s; proof 916 KiB; verification 0.16 s. Changing one output token after
proving is rejected at the output anchor's argmax constraint, and the
integer Freivalds check accepts the honest matrix product and rejects a
tampered one.

The same-trace TinyLlama experiment fixes the prompt ``The sky is'', a
maximum context of eight, and one trace and one set of weight
commitments, then proves the trace at sample settings 0, 2, 5, 10, and
all candidates. Loading and quantization took 20.6 s, weight commitments
353.2 s, and trace generation 88.2 s, all shared by every row. Each
setting seals all 47 chunk boundaries and proves the requested random
candidates plus the two anchors; the all-candidate setting runs through
the same interface with the sample clamped to the 45 candidates. Table 2
reports the results.

\begin{table*}[t]
\centering
\small
\setlength{\tabcolsep}{3.5pt}
\renewcommand{\arraystretch}{1.18}
\begin{tabularx}{\textwidth}{>{\raggedright\arraybackslash}X>{\raggedright\arraybackslash}X>{\raggedright\arraybackslash}X>{\raggedright\arraybackslash}X>{\raggedright\arraybackslash}X}
\toprule
Random samples & Proved / sealed & Seal + prove (s) & Verify (s) & Proof size \\ 
\midrule
0 & 2 / 47 & 62.5 & 0.9 & 75.50 KiB \\ 
2 & 4 / 47 & 159.1 & 1.5 & 1.25 MiB \\ 
5 & 7 / 47 & 276.8 & 1.7 & 1.77 MiB \\ 
10 & 12 / 47 & 324.7 & 3.7 & 5.08 MiB \\ 
All 45 candidates & 47 / 47 & 1,256.1 & 17.7 & 25.98 MiB \\ 
\bottomrule
\end{tabularx}
\caption{TinyLlama same-trace comparison, one observation per setting, CPU
proving and verification. Every row verified. Times include boundary
sealing and exclude the shared loading, weight-commitment, and
trace-generation stages.}
\end{table*}

At five random samples, seal-plus-prove time is \(276.8/1256.1=22.0\%\)
of full coverage, proof size \(1.77/25.98=6.8\%\), and verification 1.7
s against 17.7 s, for arithmetic coverage of seven of 47 chunks. This is
a cost comparison at different audit strengths: a fixed invalid
candidate has ideal coverage \(5/45=11.1\%\) in that row and 100\% in
the last. Anchors only is not free, because the trace is still sealed
and the endpoints proved. Cost does not grow in proportion to the number
of selected chunks either, since chunk types differ in size and each row
is one draw; a cost model would need repeated draws or a per-chunk
inventory.

After the five settings completed, the upstream monolithic
single-transcript prover was run on the same trace as an external
baseline and aborted with a stack overflow. The chunked all-candidate
row is therefore the full-coverage baseline for this implementation.

\begin{figure*}[t]
\centering
\includegraphics[width=\textwidth]{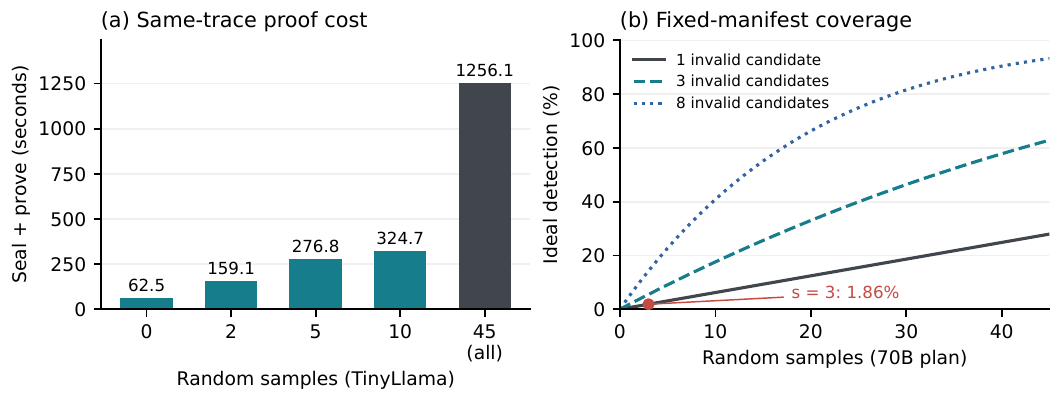}
\caption{Left: measured TinyLlama seal-plus-prove time at five audit strengths on one trace (single observations). Right: ideal detection probability for fixed invalid candidate sets under the 70B plan; the marker at three samples is the measured 70B configuration, with single-candidate coverage 1.86\%. The panels describe different models and are not a cross-model runtime comparison.}
\end{figure*}

\subsection{Packed proving and service
simulation}\label{packed-proving-and-service-simulation}

The packing sweep uses 16-token slots, up to four generated tokens per
request, and two random samples plus anchors. Table 3 and Figure 3
report the proving time of one packed proof against the cost of separate
proofs. For three requests the separate cost is measured (96.9, 105.4,
and 93.3 s); for six and twelve it is derived as \(B\times 99\) s from
the measured single-request proof.

\begin{table*}[t]
\centering
\small
\setlength{\tabcolsep}{3.5pt}
\renewcommand{\arraystretch}{1.18}
\begin{tabularx}{\textwidth}{>{\raggedright\arraybackslash}X>{\raggedright\arraybackslash}X>{\raggedright\arraybackslash}X>{\raggedright\arraybackslash}X>{\raggedright\arraybackslash}X>{\raggedright\arraybackslash}X}
\toprule
Requests \(B\) & Packed proving (s) & Separate proofs (s) & Baseline status & Proof (MiB) & Verify (s) \\ 
\midrule
1 & 99.0 & 99.0 & Single reference & Not recorded & Not recorded \\ 
3 & 116.4 & 295.6 & Measured aggregate & 2.91 & 2.5 \\ 
6 & 144.3 & 594.0 & Derived: \(6\times99\) & 3.20 & 2.9 \\ 
12 & 181.9 & 1,188.0 & Derived: \(12\times99\) & 2.11 & 2.1 \\ 
\bottomrule
\end{tabularx}
\caption{TinyLlama packing, CPU proof generation after GPU inference. Each row is
one measurement with its own selected chunks, which is why proof size
does not grow monotonically with \(B\).}
\end{table*}

Three packed requests take 39.4\% of the measured separate cost, a
factor of 2.54, and twelve take 15.3\% of the derived separate cost, a
factor of 6.53. Each additional request adds about 7 s to the packed
proof, so the per-request proving cost falls from 99 s to 15.2 s at
\(B=12\). The comparison is a cost amortization at fixed sample size,
not a speedup at equal independent-audit strength, since the twelve
requests share one sampled set (Section 4.4). The non-interference and
equivalence tests of Section 4.3 were run on the three-request layout.

\begin{figure*}[t]
\centering
\includegraphics[width=\textwidth]{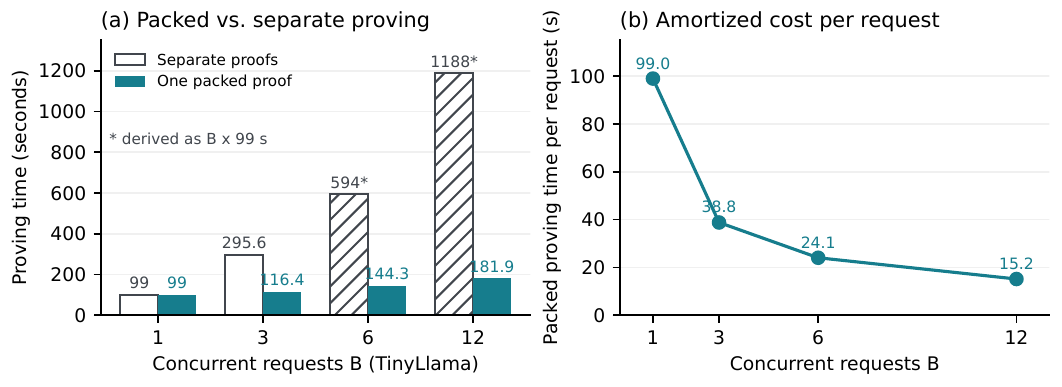}
\caption{Packed batch proving on TinyLlama. Left: one packed proof against separate proofs; hatched bars are derived as $B\times 99$ s from the measured single-request proof, the $B=3$ bar is a measured aggregate. Right: packed proving time per request. Single observations; all rows share one sampled set per batch.}
\end{figure*}

The service simulation connects registration, batched decoding, proof
serialization, and per-request binding in one flow. Twelve requests
arrive in three windows of four, six, and two; each slot has capacity 16
with at most four generated tokens, so the largest layout has six slots
and a model context of 96. Registration, which produces the
deterministic calibration, the quantized model, the weight commitments,
and a 118.14 KiB verifier context, took 417.1 s. The three windows
decoded in 101.7, 60.1, and 10.0 s, and the answers were returned before
the proofs. Proving took 118.0, 140.0, and 108.8 s (366.9 s in total,
30.6 s per request), and verification 2.2, 2.4, and 2.9 s (7.6 s in
total, 0.63 s per request). An independent verifier starts from the
distributed context and the public prompts and answers, not from the
worker's memory. In the last window the worker published a changed first
generated token after constructing the proof; the window was rejected
with a one-hot-encoding claim error, and the first two were accepted.
The per-request figures are amortizations over short requests in three
windows rather than synchronous response latencies, and the rejection
test covers post-proof answer substitution rather than a fabricated
pre-seal trace.

\subsection{A complete 70B run}\label{a-complete-70b-run}

The 70B execution uses the NousResearch distribution of Llama-2-70B with
80 transformer blocks, a maximum context of eight, three random samples
plus two anchors, and streaming weights with an 8,192 MiB store cache.
Table 4 separates registration, paid once per model version, from the
work performed for the inference. The context limit is the configured
sequence capacity, not a count of generated tokens.

\begin{table}[t]
\centering
\small
\setlength{\tabcolsep}{3.5pt}
\renewcommand{\arraystretch}{1.18}
\begin{tabularx}{\columnwidth}{>{\raggedright\arraybackslash}X>{\raggedright\arraybackslash}X>{\raggedright\arraybackslash}X}
\toprule
Stage & Time (s) & Recorded memory or output \\ 
\midrule
Load and quantize with spilling & 8,906 & About 387 GB resident peak \\ 
Create all weight commitments & 10,553 & About 60 GB working set \\ 
Registration subtotal & 19,459 & About 5.4 h \\ 
Generate integer inference trace & 48,706 & Model-sized integer weights resident \\ 
Seal 163 chunks and prove five & 1,259 & Proof: 4.34 MiB \\ 
Verify sampled proof & 46.3 & CPU verification without model weights \\ 
Per-inference subtotal & 50,011.3 & About 13.9 h \\ 
Entire recorded pipeline & 69,470.3 & About 19.3 h; completed on the 2 TB host \\ 
\bottomrule
\end{tabularx}
\caption{Llama-2-70B, one recorded execution. The stage sum is 69,470.3 s; the
program's own wall-clock total, including inter-stage overhead, was
69,500 s. No separate proving-stage memory peak was recorded.}
\end{table}

The run completed every operation needed to obtain and locally verify a
sampled proof at this scale: registration, integer inference, sealing of
all 163 boundaries, proofs of five chunks, and weight-free verification
in 46.3 s. Integer inference is 97.4\% of the per-inference time, and
the 1,259 s proof figure excludes it, so the dominant cost of a verified
70B response today is the single-threaded integer trace rather than the
proof. The streaming matters because about 2.8 TB of simultaneous
integer and field weights would exceed the host; the registration stages
ran in 387 GB and 60 GB instead. The run does not by itself give the
minimum machine size, a 70B full-coverage proof, or a 70B fidelity
score, and we do not extrapolate the five-chunk proving time to all 163
chunks or the TinyLlama GPU ratios to 70B.

\subsection{Microbenchmarks}\label{microbenchmarks}

A laptop sweep measures a synthetic feed-forward block with two linear
layers, expansion factor four, and a nonlinearity at sequence length 128
(Table 5). Proving time and peak memory grow with width across the
measured range; the local growth exponents between adjacent points vary
from 0.2 to 0.7, so the sweep supports the qualitative statement that
the working set is governed by the weight representation and does not
support a single asymptotic exponent or a per-parameter memory constant.
We no longer extrapolate it to width 8,192.

\begin{table*}[t]
\centering
\small
\setlength{\tabcolsep}{3.5pt}
\renewcommand{\arraystretch}{1.18}
\begin{tabularx}{\textwidth}{>{\raggedright\arraybackslash}X>{\raggedright\arraybackslash}X>{\raggedright\arraybackslash}X>{\raggedright\arraybackslash}X>{\raggedright\arraybackslash}X}
\toprule
Width \(d\) & Parameters (M) & Prove (s) & Peak memory (GB) & Proof (KiB) \\ 
\midrule
256 & 0.53 & 1.61 & 0.55 & 121 \\ 
512 & 2.10 & 4.45 & 2.31 & 135 \\ 
1,024 & 8.39 & 5.97 & 2.80 & 145 \\ 
2,048 & 33.6 & 9.89 & 4.52 & 156 \\ 
4,096 & 134.2 & 25.08 & 8.42 & 170 \\ 
\bottomrule
\end{tabularx}
\caption{Synthetic feed-forward measurements on Apple M3 Pro, sequence 128,
single observations.}
\end{table*}

For GPT-2 at width 768, attention-chunk proving times at sequence
lengths 16, 32, 64, 128, and 256 are 3.35, 3.66, 4.39, 5.78, and 10.05
s, against 3.94, 4.26, 4.95, 5.68, and 7.79 s for the feed-forward
chunk. The attention chunk overtakes the feed-forward chunk at sequence
256, consistent with the quadratic attention term; longer contexts were
not measured. On TinyLlama at context eight, the backend's stage timers
over one proving run attribute 40.2\% of proving time to
constraint-claim generation (the sumcheck reductions), 20.5\% to
witness-context construction, 18.5\% to proof generation, 8.8\% to
witness polynomial commitments, 6.3\% to lookup-table claims, and 5.7\%
to challenge storage. Raising the context from eight to sixteen changed
the proving time from 94.8 to 99.0 s and the proof size from 0.43 to
1.41 MiB, the observation behind packing.

\subsection{Functional negatives and
grinding}\label{functional-negatives-and-grinding}

The synthetic test matrix covers models with eight and twelve layers,
default and forced chunk counts, and sample settings from anchors only
to all candidates, and exercises proving, serialization,
deserialization, and verification. Seven negative tests each produce a
rejection: a removed chunk proof, a tampered manifest, a tampered
boundary commitment, changed public IO, verification under a different
model context, a mismatched sampling policy, and a mismatched beacon.
The beacon test verifies a proof under its original challenge and then
finds a challenge that yields a different selected set, which the
verifier rejects.

The grinding experiment fabricates a shared interior boundary in an
honest GPT-2 manifest, which invalidates the two adjacent relations, and
lets the adversary recommit that boundary until the sample avoids both
chunks. With \(M=25\) and \(b=2\), the ideal escape probability is \(q\)
and the expected number of attempts is \(1/q\). Each attempt recommits
two boundary polynomials of \(2^{14}\) elements and redraws the sample,
at about 12.5 ms on the laptop; it does not rerun the model.

\begin{table*}[t]
\centering
\small
\setlength{\tabcolsep}{3.5pt}
\renewcommand{\arraystretch}{1.18}
\begin{tabularx}{\textwidth}{>{\raggedright\arraybackslash}X>{\raggedright\arraybackslash}X>{\raggedright\arraybackslash}X>{\raggedright\arraybackslash}X>{\raggedright\arraybackslash}X>{\raggedright\arraybackslash}X}
\toprule
Samples \(s\) & Ideal detection & Expected attempts & Mean / 200 trials & Worst / 200 & Fixed-manifest escape / 2,000 challenges \\ 
\midrule
1 & 8.0\% & 1.09 & 1.11 & 3 & 91.2\% \\ 
3 & 23.0\% & 1.30 & 1.35 & 5 & 75.5\% \\ 
5 & 36.7\% & 1.58 & 1.64 & 6 & 61.7\% \\ 
10 & 65.0\% & 2.86 & 3.00 & 16 & 33.9\% \\ 
\bottomrule
\end{tabularx}
\caption{GPT-2 grinding control, two invalid candidates among 25 non-anchor
chunks, 200 trials per setting. The last column fixes the manifest and
draws the sample under 2,000 distinct challenge strings of the form
\texttt{beacon-i}; it tests the schedule, not a deployed randomness
beacon.}
\end{table*}

The measured means match \(1/q\) at the reported resolution, and the
fixed-manifest escape fractions match \(q\). Under a manifest-only
schedule the attack is an offline search whose unit cost is one boundary
recommitment; at the 70B parameters the expected search is 1.019
attempts for one invalid candidate and 1.038 for two. With the challenge
supplied after sealing, the escape probability is \(q\) per obligation,
provided the ordering and uniqueness requirements of Section 3.2 hold.
The 12.5 ms figure is a GPT-2 boundary cost measured on the laptop; the
70B values are algebraic.

\section{Canonical-Model Fidelity}\label{canonical-model-fidelity}

\subsection{Calibration identity and residual
precision}\label{calibration-identity-and-residual-precision}

Quantization produces weight values and activation scales. A seeded
calibration descriptor makes the calibration inputs reproducible, and
registration retains the descriptor, the produced artifacts, the
configuration, and the full digests; Section 7.4 measures how far
reproducibility extends to the GPU computation itself.

The initial TinyLlama failure was visible in ordinary text. The generic
observer constrained the residual stream to the nominal 12-bit setting
used for linear inputs, and greedy continuations degenerated into
fragments. The LLM-aware observer keeps the residual stream at about 24
bits and requantizes only around operations with narrower inputs.
Holding the model and prompts fixed while changing the observer produced
the largest improvement in the ablation of Table 7. Raising the global
bit width was not an equivalent remedy: a 16-bit setting exceeded an
intermediate arithmetic budget and failed, whereas the observer changes
where precision is retained. The wider residual also exposed an integer
overflow in the LayerNorm variance hint, which now is computed in
floating point; the verifier's encoded normalization constraints, with
their explicit error bounds, are unchanged.

We compared one random calibration sequence, 32 random sequences, 32
corpus sequences, and percentile-clipped corpus calibration. MinMax
tracking performed far better than 0.1/99.9-percentile clipping in this
implementation, consistent with the importance of activation outliers
reported in quantization work \citep{llmint82022, smoothquant2023}. The
result concerns the tested observer and scaling paths.

\subsection{Generation-level
diagnostic}\label{generation-level-diagnostic}

The generation diagnostic uses 16 short English prompts that were not
used for calibration, greedy decoding, eight generated tokens per
prompt, and a maximum context of 24. We report first-token agreement,
agreement at corresponding generated positions, exact agreement of the
whole continuation, and teacher-forced agreement when the canonical
model is conditioned on the f32 continuation; teacher forcing removes
the drift that follows a first differing token.

\begin{table*}[t]
\centering
\small
\setlength{\tabcolsep}{3.5pt}
\renewcommand{\arraystretch}{1.18}
\begin{tabularx}{\textwidth}{>{\raggedright\arraybackslash}X>{\raggedright\arraybackslash}X>{\raggedright\arraybackslash}X>{\raggedright\arraybackslash}X>{\raggedright\arraybackslash}X}
\toprule
Observer / calibration & First token & Generated positions & Whole continuation & Teacher forced \\ 
\midrule
Generic, one random sequence & 12\% & 5\% & 0\% & 16\% \\ 
LLM, one random sequence & 88\% & 53\% & 31\% & 89\% \\ 
LLM, 32 random sequences & 100\% & 52\% & 31\% & 87\% \\ 
LLM, 32 corpus sequences, MinMax & 88\% & 55\% & 31\% & 91\% \\ 
LLM, corpus, percentile clipping & 6\% & 1\% & 0\% & 3\% \\ 
\bottomrule
\end{tabularx}
\caption{Generation fidelity against the f32 TinyLlama execution, 16 prompts and
eight continuation tokens, integer rounding as archived. The nominal
linear-input setting is 12 bits in every row; residual handling differs
by observer.}
\end{table*}

The observer change is the central observation: teacher-forced agreement
rises from 16\% to about 90\%. With 16 prompts one success moves a
prompt-level metric by 6.25 points, and the 95\% Wilson interval of the
88\% first-token result is about 64-97\%, so the table does not rank the
three LLM-observer rows against each other. The comparison of corpus and
random calibration rests on the paired corpus evaluation below.

\subsection{Paired WikiText-2
evaluation}\label{paired-wikitext-2-evaluation}

The corpus experiment uses the WikiText-2 test split
\citep{wikitext2017} under this project's token-stream construction:
blank lines are removed, each remaining line is tokenized independently
with the model's tokenizer, and the token sequences are concatenated.
The file yields 335,634 tokens, which form 655 complete non-overlapping
windows of 512 tokens; the final 274 tokens are discarded, and each
window contributes 511 next-token targets, for 334,705 scored positions.
Both the f32 and the canonical model see exactly the same windows and
targets, which is the property the comparison needs; the absolute
perplexity is not comparable with published TinyLlama figures obtained
under other tokenization and context protocols.

At each position the script computes log-softmax of both models' logits
in f64 and records the probability of the true next token. Perplexity is
the exponential of the mean negative log-likelihood; true-token accuracy
counts whether a model's top-1 token equals the corpus target; argmax
agreement counts whether the two models choose the same token; top-5
overlap is \(|A_5\cap B_5|/5\); and KL divergence is
\(D_{\mathrm{KL}}(p_{\mathrm{f32}}\Vert p_{\mathrm{canonical}})\) in
nats, all averaged over positions. The corpus calibration uses the first
32 sequences of 512 tokens from a separate WikiText-2 training file; the
random calibration uses one seeded random sequence with seed 42. Both
use the LLM observer and MinMax tracking.

\begin{table}[t]
\centering
\small
\setlength{\tabcolsep}{3.5pt}
\renewcommand{\arraystretch}{1.18}
\begin{tabularx}{\columnwidth}{>{\raggedright\arraybackslash}X>{\raggedright\arraybackslash}X>{\raggedright\arraybackslash}X>{\raggedright\arraybackslash}X}
\toprule
Metric & f32 reference & Corpus calibration & Random calibration \\ 
\midrule
Perplexity & 11.94 & 12.95 & 12.94 \\ 
True-token top-1 accuracy & 50.5\% & 49.2\% & 49.2\% \\ 
Argmax agreement with f32 & 100\% & 84.8\% & 84.9\% \\ 
Mean top-5 overlap & 100\% & 82.5\% & 82.5\% \\ 
Mean KL, nats / position & 0 & 0.1182 & 0.1182 \\ 
\bottomrule
\end{tabularx}
\caption{Paired WikiText-2 test evaluation, 655 windows and 334,705 positions.
Corpus calibration uses 32 sequences; random calibration uses one seeded
sequence.}
\end{table}

The two arms are indistinguishable at the logged precision: argmax
agreement differs by 0.1 percentage points and mean KL agrees to four
decimals, so the choice between these two calibration descriptors did
not affect fidelity in this experiment. Perplexity rises by about 8.5\%
and true-token accuracy falls by 1.3 points; argmax disagreement is
about 15\%. The first pair measures language-model quality under the
scoring protocol, the second disagreement with this particular
reference; neither converts directly into a probability that a whole
generated response differs. Adjacent positions are correlated and the
aggregate logs do not retain per-window statistics, so we report totals
rather than a confidence interval.

\begin{figure*}[t]
\centering
\includegraphics[width=\textwidth]{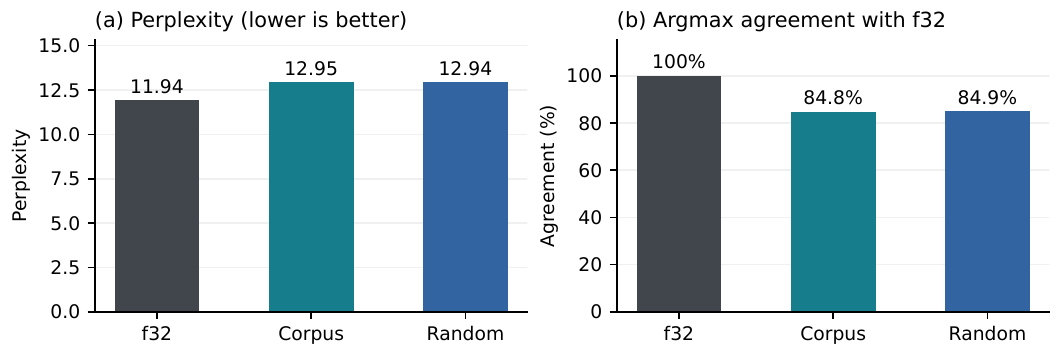}
\caption{Paired TinyLlama evaluation on the WikiText-2 test split. Perplexity and top-1 agreement measure different properties; the two calibration arms coincide at logged precision.}
\end{figure*}

\subsection{Repeated GPU execution}\label{repeated-gpu-execution}

Ten separate registrations of the same seeded configuration printed
identical calibration, linear-weight, and activation-scale fingerprint
prefixes (12 hexadecimal characters each). Within each registration, two
repeated inferences on the same input were compared with the first: 16
of 20 matched token for token and four differed, once in run five, twice
in run eight, and once in run nine. The registered model version
therefore reproduces across processes, while the GPU inference of a
fixed model on a fixed input still varies. The variability lies in
floating-point execution order on the device rather than in calibration;
identifying the kernel is future work. An earlier four-registration
observation of one differing fingerprint did not retain comparable data
and is kept as history rather than combined into a 14-trial statistic.

\subsection{Scope of the fidelity
result}\label{scope-of-the-fidelity-result}

The evaluation compares the canonical model with an f32 execution of the
same checkpoint in the same framework. It contains no matched fp16,
W8A8, GPTQ, or AWQ baseline on the same windows
\citep{gptq2023, awq2024}, and those methods quantize different
quantities under different constraints, so the 85\% agreement is a
measured gap for this canonical model rather than a ranking against
conventional low-bit inference. Fidelity for the 70B model and for
packed slots at nonzero offsets has not been measured, and a registered
layout needs its own report. A published fidelity report identifies
itself by digest; whether it was honestly produced is a governance
question that an auditor answers by reproducing it on public artifacts.

\section{Deployment Implications and
Limitations}\label{deployment-implications-and-limitations}

\subsection{Coverage, cheating cost, and
enforcement}\label{coverage-cheating-cost-and-enforcement}

An audit budget fixes a sample size and hence a conditional detection
probability; a mechanism must turn that probability into a deterrent.
For a risk-neutral worker let \(g\) be the gain from cheating on one
obligation, \(a\) the probability that the obligation is audited, \(p\)
the conditional detection probability, and \(F\) the enforceable
penalty. A necessary condition is \(apF>g\). At the
single-invalid-candidate 70B setting \(p=3/161\), so \(F/g>53.7\) if
every obligation is audited and \(F/g>537\) at a 10\% audit rate. These
figures show what weak single-candidate coverage demands of the penalty
layer; they are illustrations, not a deposit schedule. The condition is
void if the registrar is controlled by the worker, the beacon is
selectable, deadlines are unenforced, or several manifests are accepted
for one obligation. Cheating gain also depends on the attack:
fabricating one isolated boundary invalidates two relations and saves
little compute, replacing a long suffix changes many relations, and
algebraically structured alternative weights might satisfy every
relation at lower cost. A practical cost model needs end-to-end attacks,
not only the hit probability.

\subsection{Approximate relations and model
authenticity}\label{approximate-relations-and-model-authenticity}

Layerwise tolerance checks do not in general compose into network-wide
authenticity: Zamir constructs functionally equivalent ReLU networks in
which small layerwise deviations steer the final result
\citep{zamir2026}. SLP's coverage bound concerns false encoded relations
and gives no guarantee against a trace that differs from the reference
execution while satisfying every relation. The normalization gadgets of
Section 2.1 admit an approximation range, and establishing that every
witness in that range yields the same tokens, or bounding the
network-wide effect, is open; full chunk coverage verifies the encoded
relation, not that equivalence. Hollow-LLM shows a separate gap between
satisfying a model's equations and expending the advertised computation
under committed private weights \citep{hollowllm2026}. Authenticating
public weights limits substitution of a hollow model and does not turn a
correctness proof into a lower bound on effort, so we make no
compute-forcing claim at any coverage. Native fp16 serving followed by a
canonical replay would keep the integer runtime off the serving path; a
replay proof still refers to the replayed relation, and acceptance
thresholds would need matched benign and adversarial evaluation. We
leave that mode as an open design question.

\subsection{Scalability and empirical
validity}\label{scalability-and-empirical-validity}

The largest model result is one short-context CPU run, and the largest
GPU model is TinyLlama-1.1B, whose scoring experiment reaches 512-token
windows on a 24 GB device; these do not combine into 70B GPU serving at
context 512. Integer weight residency at 64 bits, trace memory, and
unoptimized kernels remain the major costs, the full 70B relation was
not proved, and no 70B fidelity or standard-task score exists. Most
timings are single observations without dispersion, some historical
machine and dependency details were not recorded, and different sample
settings select different chunks. The service experiment uses short
requests and includes a deliberate rejection; the six- and
twelve-request separate-proof baselines are derived from one measured
single-request proof. Memory figures have three different meanings in
this paper: a resident peak during loading, a working set during
commitment, and completion on a host of stated size. No proving-stage
peak was logged for 70B, no memory-versus-sample curve was measured, and
inference does not release weights layer by layer. Mixture-of-experts
routing and proof aggregation across service windows are outside the
evaluated implementation.

\subsection{Registration and artifact
boundaries}\label{registration-and-artifact-boundaries}

A deployed registry must authenticate the full model configuration, the
verification context, a compatible reference string, the calibration
artifacts, preprocessing, decoding policy, and any layout-specific
fidelity report; weight and scale hashes are necessary and not
sufficient. The library does not yet enforce a global request namespace
or seal deadlines. The local fork's root license and package-license
declarations are inconsistent, so this manuscript confers no permission
to redistribute the modified backend; the source package contains the
paper, bibliography, derived tables, and figure scripts, and a release
of the engine requires a separately resolved licensing plan.

\section{Related Work}\label{related-work}

\subsection{Full and decomposed arithmetic
proofs}\label{full-and-decomposed-arithmetic-proofs}

zkLLM develops proof techniques for large transformer computations,
including their nonlinear and attention operations \citep{zkllm2024},
and DeepProve proves complete generated sequences through its graph and
proof pipeline \citep{deepprove2026}. These systems supply the
arithmetic arguments; SLP changes which chunks are proved and how weight
data is held, and does not improve their full-coverage bounds.
zkComposer partitions proof construction and links adjacent partitions
through shared activation commitments \citep{zkcomposer2026}; it proves
partitions in parallel for time and sequentially for memory, reporting
up to 8.1 times lower peak memory on GPT-2 in sequential mode. Holding
only the partition being proved is therefore established; SLP's systems
contribution is disk-backed integer weights and streamed weight
commitments in this backend, combined with sampled proving and exercised
at 70B. NanoZK chains layerwise proofs and triages an audit budget with
Fisher information, requiring all layers for full soundness
\citep{nanozk2026}. SLP shares the separation of coverage from per-proof
strength and differs in its backend, its IO-derived anchors, its packed
traces, and its registration path; it does not provide NanoZK's privacy
properties or weighted triage.

\subsection{Audits and approximate
verification}\label{audits-and-approximate-verification}

TensorCommitments studies lightweight commitments and partial checks for
language-model inference \citep{tensorcommitments2026}; as in SLP, a
committed trace can hold invalid computation outside the checked region,
and a selected opening and a selected arithmetic proof differ in cost
and in what they establish. IMMACULATE audits selected requests through
verifiable computation \citep{immaculate2026}; request sampling and
within-request chunk sampling are different budgets that a system may
combine, with packing adding a dependence because several requests share
one sampled set. TAO verifies floating-point operators with tolerance
awareness \citep{tao2026}, and approximate sumcheck gives sound
protocols for approximate computations \citep{bitan2026}; both show why
the accepted relation must be specified exactly, and neither supplies a
composition theorem for arbitrary local tolerances across a transformer
graph.

\subsection{Quantization and registered
behavior}\label{quantization-and-registered-behavior}

LLM.int8 and SmoothQuant address activation outliers in quantized
execution \citep{llmint82022, smoothquant2023}, and GPTQ and AWQ study
low-bit weight quantization \citep{gptq2023, awq2024}. Their
representations differ from the proof backend's fixed-point constraints,
and they are the reference methods for a future matched fidelity
comparison. The empirical contribution here is a failure diagnosis
inside a verifiable-inference pipeline: the observer's treatment of the
residual stream, not the calibration data, determined whether the
canonical model tracked its parent. Registration should publish both the
computation and the evidence about its behavior, and keep the two
assurances distinct: a commitment does not establish quality, and a good
perplexity does not establish identity.

\section{Conclusion}\label{conclusion}

SLP turns audit coverage into a runtime parameter over one set of
boundary commitments and shows what that parameter costs. On one
TinyLlama trace, proving seven of 47 chunks takes 22\% of the time and
7\% of the proof size of proving all 47. Packing twelve concurrent
requests into one trace proves them in 181.9 s, 6.5 times less than
separate proofs at the measured single-proof cost, with each request's
prompt and answer bound to its slot and a tampered answer rejected in a
simulated service. Streamed registration completes a Llama-2-70B
sealing, sampled proof, and 46 s weight-free verification on a 2 TB
host. The canonical model's fidelity, after the observer repair, is
84.8-84.9\% argmax agreement with the floating-point reference over
334,705 positions.

The limits are equally concrete. Commitment coverage exceeds arithmetic
coverage, and a fixed invalid chunk in the measured 70B setting is
covered with probability 3/161. A manifest-only Fiat-Shamir schedule is
grindable at millisecond cost and requires an externally ordered
challenge with a unique accepted manifest per obligation. The
measurements use a test reference string, and exact verification of a
bounded-error relation does not establish unique native-model execution.
A service that supplies authenticated registration, a proper setup,
pre-challenge manifest uniqueness, and an audited encoding can use the
demonstrated engineering with the guarantees stated here.

\section*{Declarations}\label{declarations}
\addcontentsline{toc}{section}{Declarations}

\textbf{Data and code availability.} The raw run logs, extracted data
tables, and the experiment report are publicly available at
https://github.com/TrueOpen/slp-experiments. The manuscript source
package includes the derived numerical tables and the scripts for the
figures. The checkpoint families and corpus are identified in the text.
The modified backend is not included because its redistribution status
is unresolved.

\textbf{Ethics.} The work consists of software experiments on existing
model checkpoints, public corpus material, and short diagnostic prompts.
It involves no human participants and no newly collected personal data.
The adversarial examples are controlled local verification tests, not
attacks on third-party services.

\textbf{AI assistance.} Claude Code and Codex assisted implementation,
analysis, drafting, reference checks, and manuscript preparation.
AI-generated review reports were used as internal revision aids; their
recommendations were checked against code, logs, and source
publications, and they are not independent human peer review.

\textbf{Authorship, funding, and competing interests.} Both authors are
with TrueOpen. This work was carried out and funded by TrueOpen.
TrueOpen is developing an open inference network that intends to deploy
the protocol described here; the authors therefore have a commercial
interest in its adoption. Correspondence: https://github.com/TrueOpen.
\bibliographystyle{plainnat}
\bibliography{references}
\end{document}